\documentclass[12pt]{iopart}

\usepackage{dcolumn} 
\usepackage{bm} 
\usepackage{graphicx}
\usepackage{color}

\def\pt{\mbox{$p_{\mathrm t} $}}

\def\snn{\mbox{$\sqrt{s_{_{\rm NN}}}$}}   

\newcommand{ \be }{\begin{eqnarray}}
\newcommand{ \ee }{\end{eqnarray}}

\definecolor{dgreen}{cmyk}{1.,0.,1.,0.4}        
\definecolor{orange}{cmyk}{0.,0.353,1.,0.}    

\begin{document}
\today

\title{Elliptic Flow: A Brief Review}
\author{Raimond Snellings}
\address{Utrecht University, P.O. Box 80000, 3508 TA Utrecht, 
The Netherlands}

\begin{abstract}
One of the fundamental questions in the field of subatomic physics  is what happens to matter at extreme densities and temperatures 
as may have existed in the first microseconds after the Big Bang and exists, perhaps, in the core of dense neutron stars.
The aim of heavy-ion physics is to collide nuclei at very high energies and thereby create such a state of matter in
the laboratory.
The experimental program started in the 1990's with collisions made available at 
the Brookhaven Alternating Gradient Synchrotron (AGS), the
CERN Super Proton Synchrotron (SPS) and continued at the 
Brookhaven Relativistic Heavy-Ion Collider (RHIC) with maximum 
center of mass energies of 
$\sqrt{s_{_{\rm NN}}}$ = 4.75, 
17.2 and 200~GeV respectively.
Collisions of heavy-ions at the unprecedented energy of 2.76~TeV have recently become available at 
the LHC collider at CERN.
In this review I will give a brief introduction to the physics of ultra-relativistic
heavy-ion collisions and discuss the current status of elliptic flow measurements.
\end{abstract}

\maketitle

\section{Heavy-Ion Physics}
To our current understanding, the universe went
through a series of phase transitions which mark the most important epochs of the
expanding universe after the Big Bang.  
At $10^{-11}$ s and at a temperature of 
$T \sim$ 100~GeV ($\sim 10^{15}$~K) the electroweak phase
transition took place where most of the known elementary particles
acquired their Higgs masses~\cite{Higgs:1964ia, Higgs:1964pj,
  Englert:1964et}.  
At $10^{-5}$ s and at a temperature of 
$\sim 200$~MeV ($\sim 10^{12}$~K), 
the strong phase transition took place where 
the quarks and gluons became
confined into hadrons and where the approximate chiral symmetry was spontaneously
broken~\cite{Nambu:1961tp}.

Quantum Chromodynamics ({\sc QCD}) is the underlying theory of the strong force. 
Although its fundamental degrees of freedom (quarks and
gluons) cannot be observed as free particles the {\sc QCD} Lagrangian is well established.
One of the key features of {\sc QCD} is the self coupling of the gauge bosons (gluons) which cause the 
coupling constant to increase with decreasing momentum transfer.
This running of the coupling constant gives rise to asymptotic freedom~\cite{Gross:1973ju,Politzer:1973fx} and confinement at large 
and small momentum transfers, respectively.
At small momentum transfer nonperturbative corrections, which are notoriously hard to calculate, become important.
For this reason two important nonperturbative properties of QCD, confinement and chiral symmetry breaking, 
are still poorly understood from first principles. 

One of the fundamental questions in QCD phenomenology is what the
properties of matter are at the extreme densities and temperatures where the quarks
and gluons are in a deconfined state, the so-called Quark Gluon Plasma ({\sc QGP})~\cite{Shuryak:1980tp}.
Basic arguments~\cite{Shuryak:1980tp} allow us to estimate the energy 
density $\epsilon \sim$~1~GeV/fm$^3$ and temperature 
$T \sim$~200~MeV at which the strong phase transition takes place. 
These values imply that 
the transition occurs in a regime where the coupling constant is 
large so that we can not rely anymore on perturbative {\sc QCD}.
Better understanding of the non-perturbative domain comes from
lattice {\sc QCD}, where the field equations are solved 
numerically on a discrete space-time grid.
Lattice {\sc QCD} 
provides quantitative information on the {\sc QCD}
phase transition and the Equation of State ({\sc EoS}) of the deconfined state.
At extreme temperatures (large momenta) we expect that the quarks and gluons are 
weakly interacting and that the {\sc QGP} would behave as an ideal gas.
For an ideal massless gas the {\sc EoS} is given by:
\begin{eqnarray}\label{eq:EoS}
P & = & \frac{1}{3} \epsilon, 
\;\;\;
\epsilon = g \frac{\pi^2}{30} T^4,
\end{eqnarray}
where $P$ is the pressure, 
$\epsilon$ the energy density, $T$ 
the temperature and $g$ is the effective number of degrees of freedom. 
Each bosonic degree of freedom contributes 
1 unit to $g$, whereas each fermionic
degree of freedom contributes $\frac{7}{8}$. 
The value of $g = 47.5$ for a three flavor QGP which is an order of magnitude
larger than that of a pion gas where $g \sim 3$. 
  
\begin{figure}[hbt]
  {\center
    \includegraphics[width=0.65\textwidth]{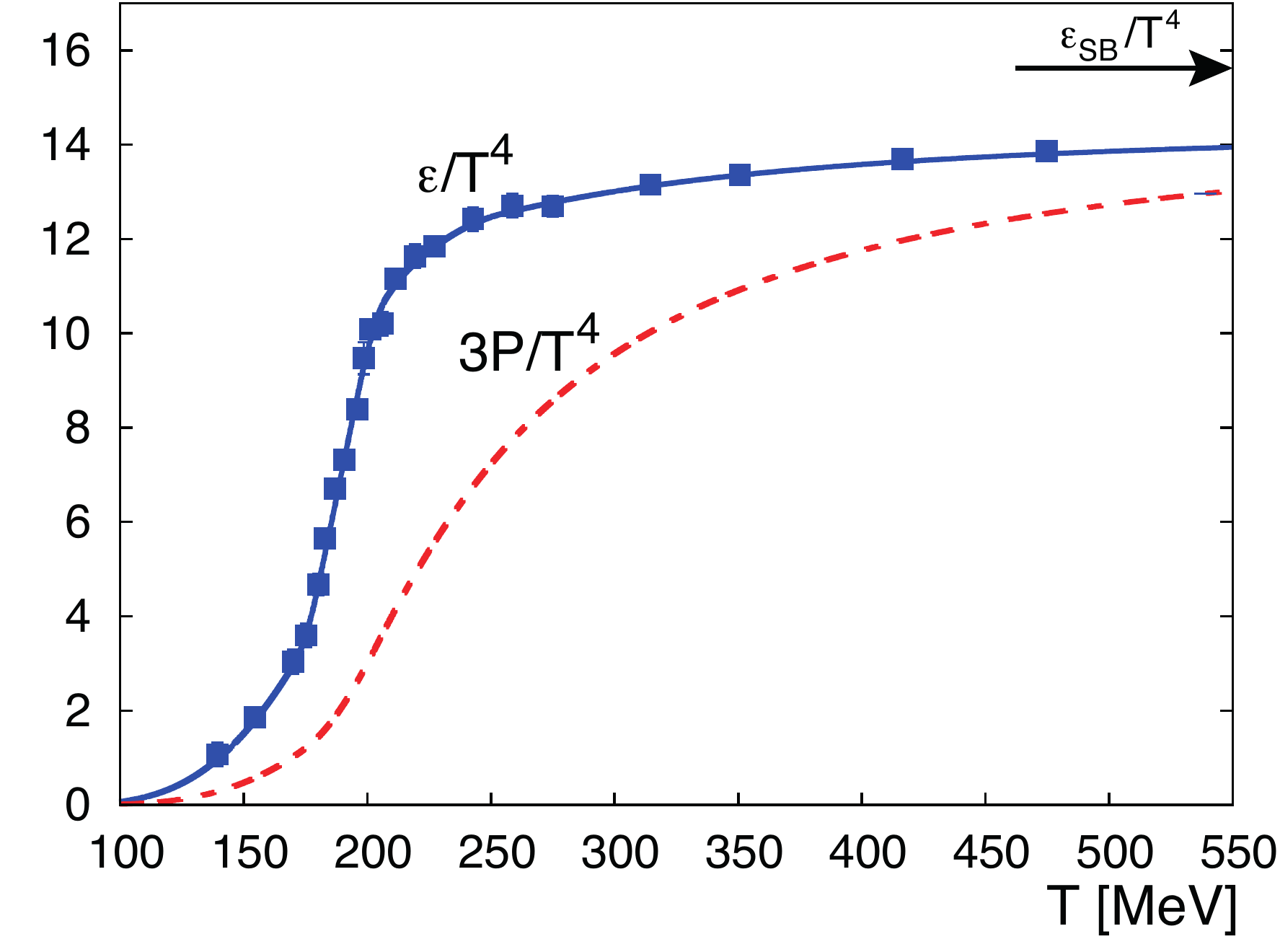}
    \caption{
      Energy density $\epsilon/T^4$ (full curve) and pressure $3P/T^4$ (dashed curve) as a function of 
      temperature $T$ from lattice calculations~\cite{Bazavov:2009zn}. 
      The arrow indicates the Stefan Boltzmann limit of the energy density.
    }
    \label{LQCD}
  }
\end{figure}
Figure~\ref{LQCD} shows the temperature dependence of the energy density as calculated from 
lattice~QCD~\cite{Bazavov:2009zn}. 
It is seen that the energy density changes rapidly around 
$T \sim$~190 MeV, which is 
due to the rapid increase in the effective degrees of freedom (lattice calculations show that the transition is a crossover).
Also shown in Fig.~\ref{LQCD} is the pressure which changes slowly compared to the 
rapid increase of the energy density around $T = 190$ MeV. 
It follows that the speed of sound, $ c_{\rm s } = \sqrt{\partial P/\partial \epsilon}$, is
reduced during the strong phase transition.
At large temperature the energy density reaches
a significant fraction ($\sim 0.9$) of the ideal massless gas limit (Stefan-Boltzmann limit).

Relativistic heavy-ion collisions are a unique
tool to create and study hot {\sc QCD} matter and its phase transition under controlled 
conditions~\cite{Shuryak:1980tp, Chapline:1974zf, Lee:1974ma, Lee:1978mf,
Collins:1974ky, Pisarski:1983ms}.  
As in the early universe, the hot and dense system created in a 
heavy-ion collision will expand and cool down. 
During this evolution the system probes a range of energy densities and temperatures, 
and possibly different phases. 
Provided that the quarks and gluons undergo multiple interactions
the system will thermalize and form the QGP which subsequently undergoes  a collective expansion 
and eventually becomes so dilute that it hadronizes. 
This collective expansion is called flow.

Flow is an observable that provides experimental information on the equation of state 
and the transport properties of the created {\sc QGP}. 
The azimuthal anisotropy in particle production is the clearest experimental signature of 
collective flow in heavy-ion collisions~\cite{Ollitrault:1992bk,Voloshin:2008dg,Heinz:2009xj,Huovinen:2006jp,Teaney:2009qa} .
This so-called anisotropic flow is caused by the initial asymmetries in the geometry of
the system produced in a non-central collision. 
The second Fourier coefficient of the azimuthal asymmetry is called elliptic flow.
In this report I will describe the relation between elliptic flow and the geometry of the collision
(section~\ref{eventchar} and~\ref{anflow}), the sensitivity of elliptic flow to the EoS and transport properties (section~\ref{anflow})
and the techniques used to measure elliptic flow from the data (section~\ref{methods}).
In section~\ref{measurements} I will review the elliptic flow measurements at the LHC and at lower energies, together with
the current theoretical understanding of these results.    

\section{Event Characterization}
\label{eventchar}

\begin{figure}[htb]
  \begin{center}
    \includegraphics[width=1.\textwidth]{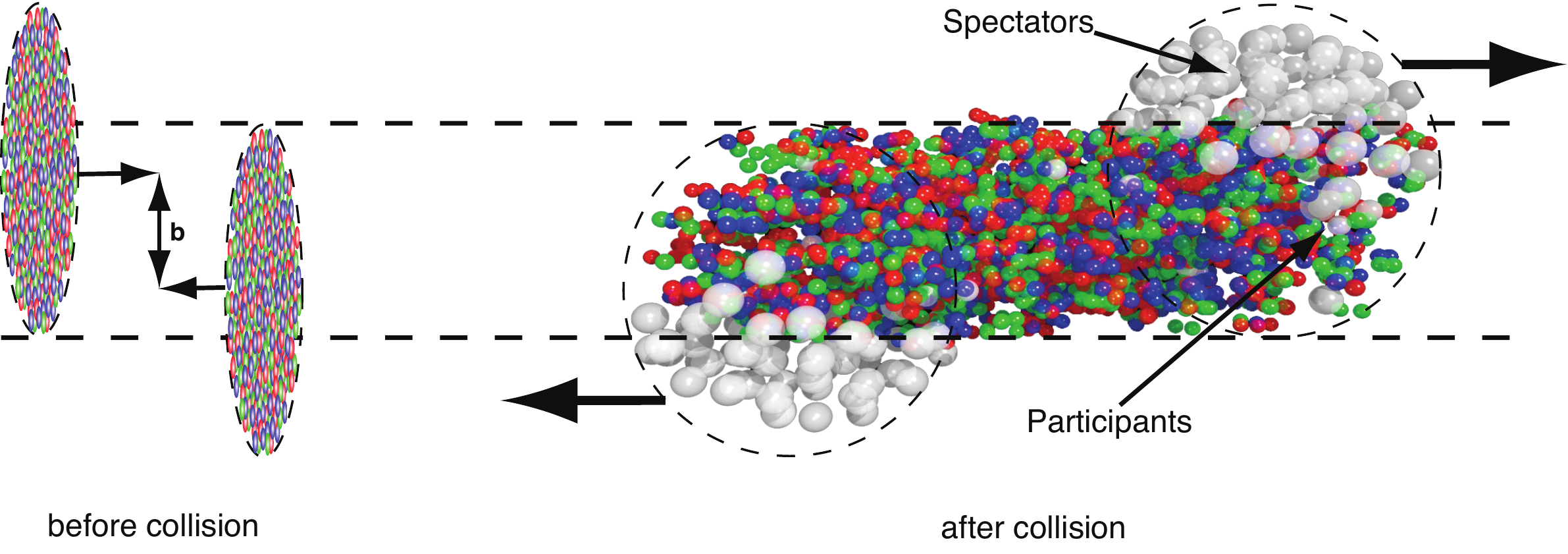}
    \caption{
      Left: The two heavy-ions before collision with impact parameter {\bf b}.
      Right:  The spectators continue unaffected, while in the participant zone particle production takes place.
      \label{centrality} }
  \end{center}
\end{figure}
Heavy-ions are extended objects and the system created in a head-on collision is different from that in a 
peripheral collision. 
To study the properties of the created system, collisions are therefore categorized by their centrality. 
Theoretically the centrality is 
defined by the impact parameter ${\bf b}$ (see Fig.~\ref{centrality}) which, 
however, cannot be directly observed. 
\begin{figure}[htb]
    \begin{center}
      \includegraphics[width=\textwidth]{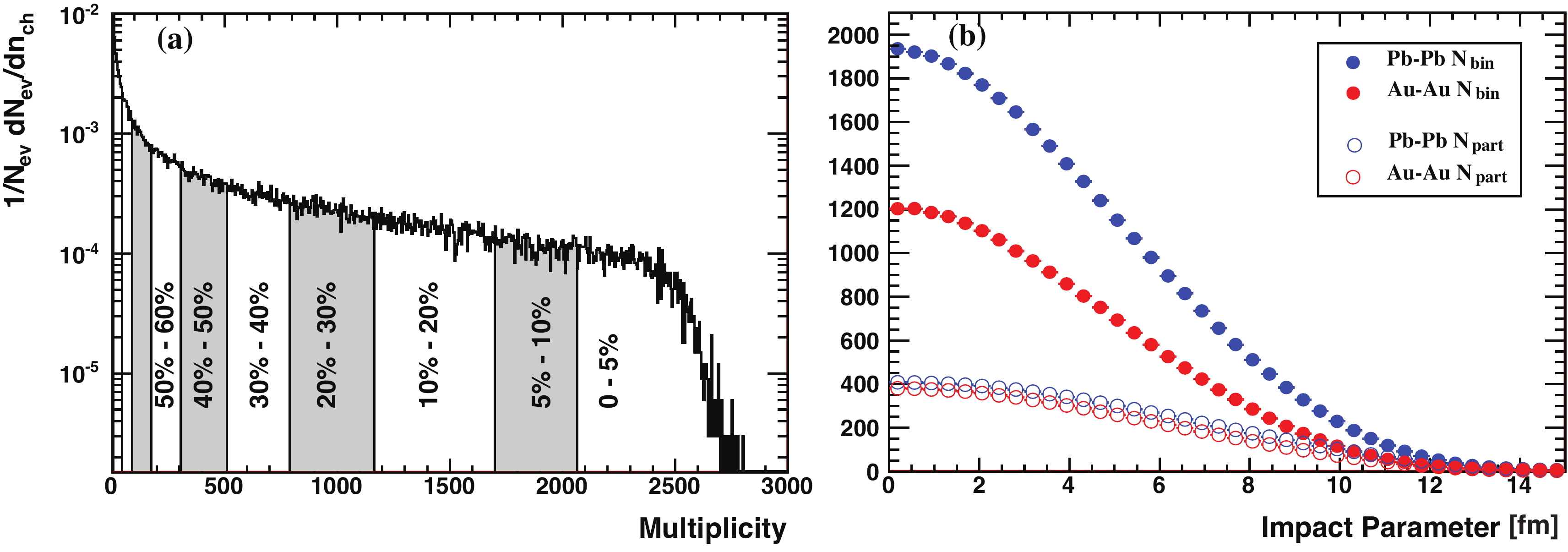}
      \caption{ 
        a) Charged particle distribution from Pb-Pb collisions at \snn~=~2.76~TeV measured with ALICE, 
        showing a classification in centrality percentiles (from~\cite{Aamodt:2010pa}).
        b) Number of participating nucleons $N_{\rm part}$ and binary collisions $N_{\rm bin}$
        versus impact parameter for Pb-Pb and Au-Au collisions at \snn~=~2.76 and 0.2~TeV, respectively.
      \label{alicemult} }
    \end{center}
\end{figure}
Experimentally, the collision centrality can be inferred 
from the measured particle multiplicities, given the assumption that the
multiplicity is a monotonic function of ${\bf b}$. 
The centrality is then characterized by the fraction, $\pi b^2/\pi(2 R_{A})^2$,  
of the geometrical cross-section 
with $R_{A}$ the nuclear radius (see Fig.~\ref{alicemult}a). 

Instead of by impact parameter, the centrality is also often characterized by
the number of participating nucleons (nucleons that undergo at least one inelastic collision) 
or by the number of equivalent binary collisions. 
Phenomenologically
it is found that the total particle production scales with the number of
participating nucleons whereas hard processes scale with the number of
binary collisions. 
These measures can be related to the impact parameter ${\bf b}$ using a
realistic description of the nuclear geometry in a Glauber
calculation~\cite{Miller:2007ri}, as is shown in Fig.~\ref{alicemult}b. 
This Figure also shows that Pb--Pb collisions at \snn~=~2.76~TeV and Au-Au at \snn~=~0.2~TeV have 
a similar distribution of participating nucleons. The number of binary collisions increases from Au--Au to Pb--Pb by about 50\% because 
the nucleon-nucleon inelastic cross section increases by about that amount at the respective center of mass energies of 0.2 and 2.76~TeV.

\section{Anisotropic Flow}
\label{anflow}

Flow signals the presence of multiple interactions between the constituents of the medium created in the collision. 
More interactions usually leads to a larger magnitude of the flow 
and brings the system closer to thermalization. 
The magnitude of the flow is therefore a detailed probe of the level of thermalization. 
\begin{figure}[htb]
  \begin{center}
    \includegraphics[width=0.80\textwidth]{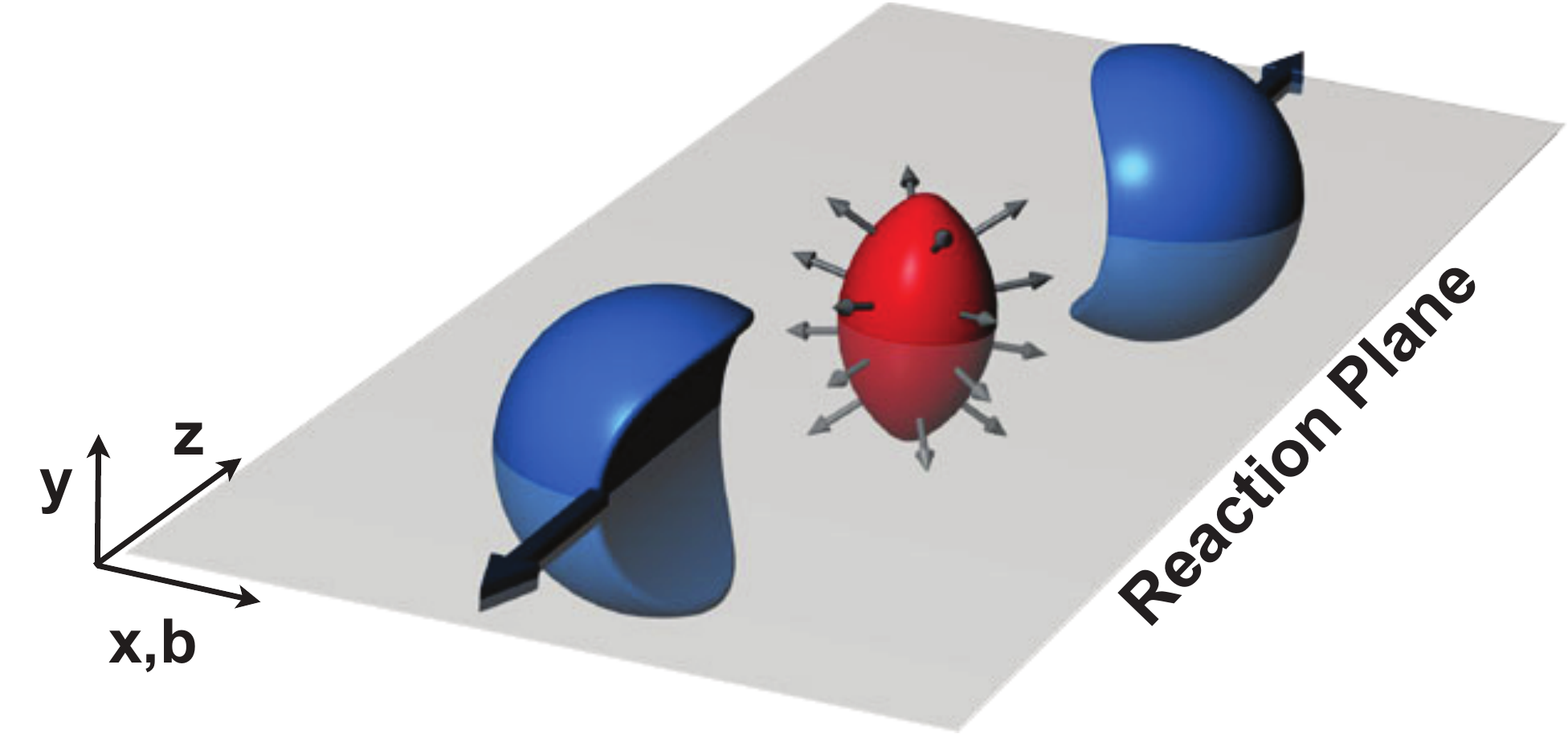}
    \caption{Almond shaped interaction volume after a non-central collision of two nuclei. The spatial anisotropy
    with respect to the $x$-$z$ plane (reaction plane) translates into a momentum anisotropy of the produced particles (anisotropic flow).}
    \label{flow_illustration}
  \end{center}
\end{figure}
The theoretical tools to describe flow are
hydrodynamics or microscopic transport (cascade) models. 
In the transport models flow depends on the opacity of the medium, be it
partonic or hadronic.  Hydrodynamics becomes applicable 
when the mean free path of the particles is much smaller than the system
size, and allows for a description of the system in terms of macroscopic
quantities. This gives a handle on the equation of state of the
flowing matter and, in particular, on the value of the sound
velocity $c_{s}$. 

Experimentally, the most direct evidence of flow comes from the observation 
of anisotropic flow which is the anisotropy in particle momentum distributions correlated 
with the reaction plane. 
The reaction plane is defined by the impact parameter and the beam direction $z$
(see Fig.~\ref{flow_illustration}).
A convenient way of characterizing the various patterns of anisotropic
flow is to use a Fourier expansion of the invariant triple differential distributions:
\begin{equation}
E \frac{{\mathrm d}^3 N}{{\mathrm d}^3 {\bf p}} = 
\frac{1}{2\pi} \frac {{\mathrm d}^2N}{p_{\mathrm t}{\mathrm d}p_{\mathrm t}{\mathrm d} y} 
  \left(1 + 2\sum_{n=1}^{\infty}  v_n \cos[n(\varphi-\Psi_{\rm RP})]\right), 
\label{invariantyield}
\end{equation}
where $E$ is the energy of the particle, $p$ the momentum, $p_{\rm t}$
the transverse momentum, $\varphi$ the azimuthal angle, $y$ the rapidity,
and $\Psi_{\rm RP}$ the reaction plane angle.
The sine terms in such an
expansion vanish because of the reflection symmetry with respect to the
reaction plane. The Fourier coefficients are $\pt$ and $y$ dependent and are given by
\begin{equation}
v_n(\pt,y) = \langle \cos[n(\varphi-\Psi_{\rm RP})] \rangle,
\label{fouriercoeff}
\end{equation}
where the angular brackets denote an average over the 
particles, summed over all events, in the $(\pt,y)$ bin under study.  
In this Fourier decomposition, the coefficients $v_1$ and $v_2$ 
are known as directed and elliptic flow, respectively.

\begin{figure}[tb]
  {\center
    \includegraphics[width=1.\textwidth]{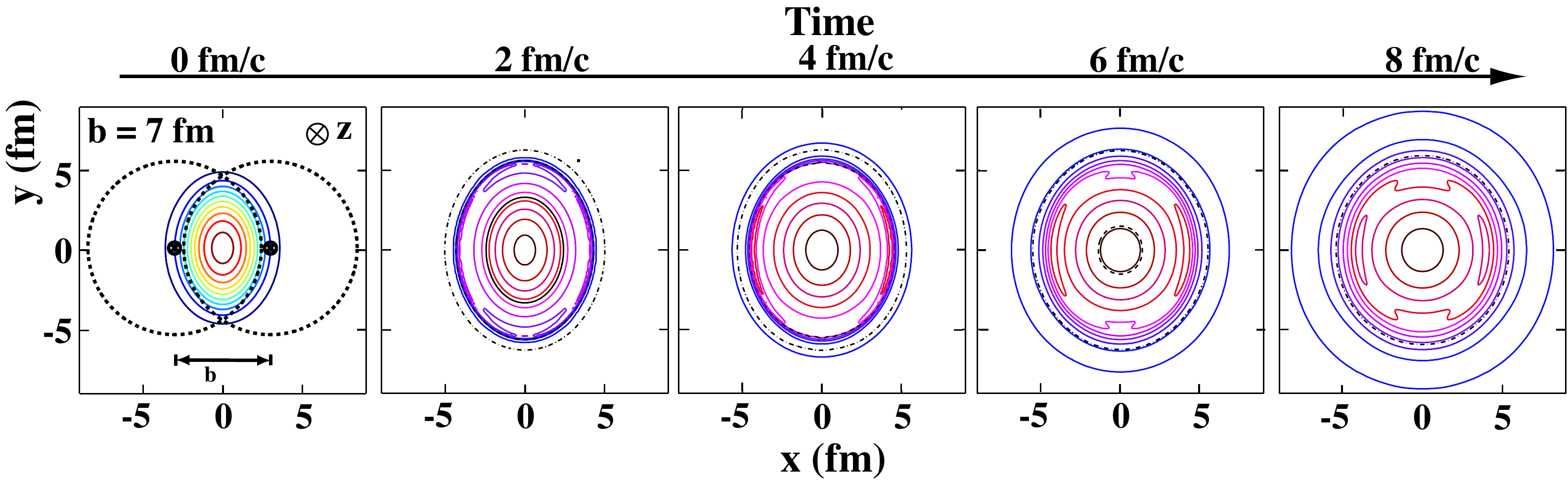}
    \caption{
      The created initial transverse energy density
      profile and its time dependence in coordinate
      space for a non-central heavy-ion collision~\cite{Kolb:2003dz}. The $z$-axis 
      is along the colliding beams, the $x$-axis is defined by the
      impact parameter.
    }
    \label{hydroinitial}
  }
\end{figure}
The evolution of  the almond shaped interaction volume 
is shown in Fig.~\ref{hydroinitial}.
The contours indicate the energy density profile and the plots from left to right show 
how the system evolves from an almond shaped transverse overlap region into an 
almost symmetric system. During this expansion, governed by the velocity of 
sound, the created hot and dense system cools down. 
 
\begin{figure}[th]
  {\center
      \includegraphics[width=1.\textwidth]{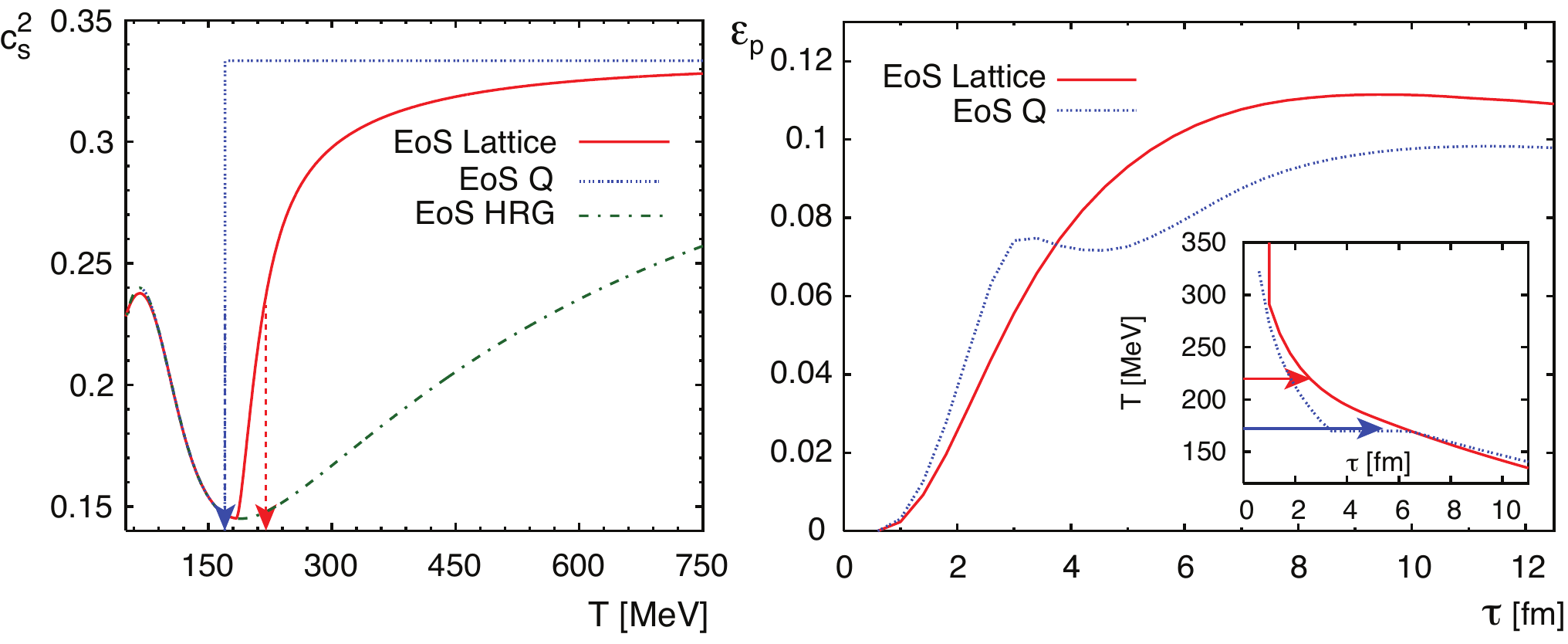}
    \caption{
    a) The velocity of sound squared versus temperature for three equations of state~\cite{Huovinen:2009yb}.
    b) The anisotropy in momentum space for two equations of state used in hydrodynamic calculations~\cite{Huovinen:2009yb}.
    }
    \label{EoShydro}
  }
\end{figure}
Figure~\ref{EoShydro}a shows the velocity of sound versus temperature for three different equations of state~\cite{Huovinen:2009yb}. 
The dash-dotted line is the hadron resonance gas EoS, 
the red full line is a parameterization of the EoS which matches recent lattice calculations and the blue dashed line 
is an EoS which incorporates a first order phase transition.  The arrows indicate the corresponding 
transition temperatures for the lattice inspired EoS and the EoS with a first order phase transition.
The temperature dependence of the sound velocity clearly differs significantly between the different equations of state.
Because the expansion of the system and the buildup of collective motion depend on the velocity of sound, it is expected that this 
difference will have a clear signature in the flow.
The buildup of the flow for two different EoS is shown in Fig.~\ref{EoShydro}b. 
Due to the stronger expansion in the reaction plane the initial almond shape anisotropy in coordinate space vanishes, 
as was shown in Fig.~\ref{hydroinitial}, while the momentum space distribution changes in the opposite direction 
from being approximately azimuthally symmetric to having a preferred 
direction in the reaction plane. The asymmetry in momentum space can be quantified by:
\begin{equation}
\varepsilon_{p} = \frac{\langle T_{xx} - T_{yy} \rangle}{\langle T_{xx} + T_{yy} \rangle},
\label{eq:epsilon_p}
\end{equation}
where $T_{xx}$ and $T_{yy}$ are the diagonal transverse components of the energy momentum tensor and the brackets denote an averaging 
over the transverse plane.
Figure~\ref{EoShydro}b shows that $\varepsilon_{p}$ versus time starts at zero after which the anisotropy quickly develops and is indeed dependent 
on the EoS.  

\begin{figure}[bth]
  {\center
      \includegraphics[width=1.\textwidth]{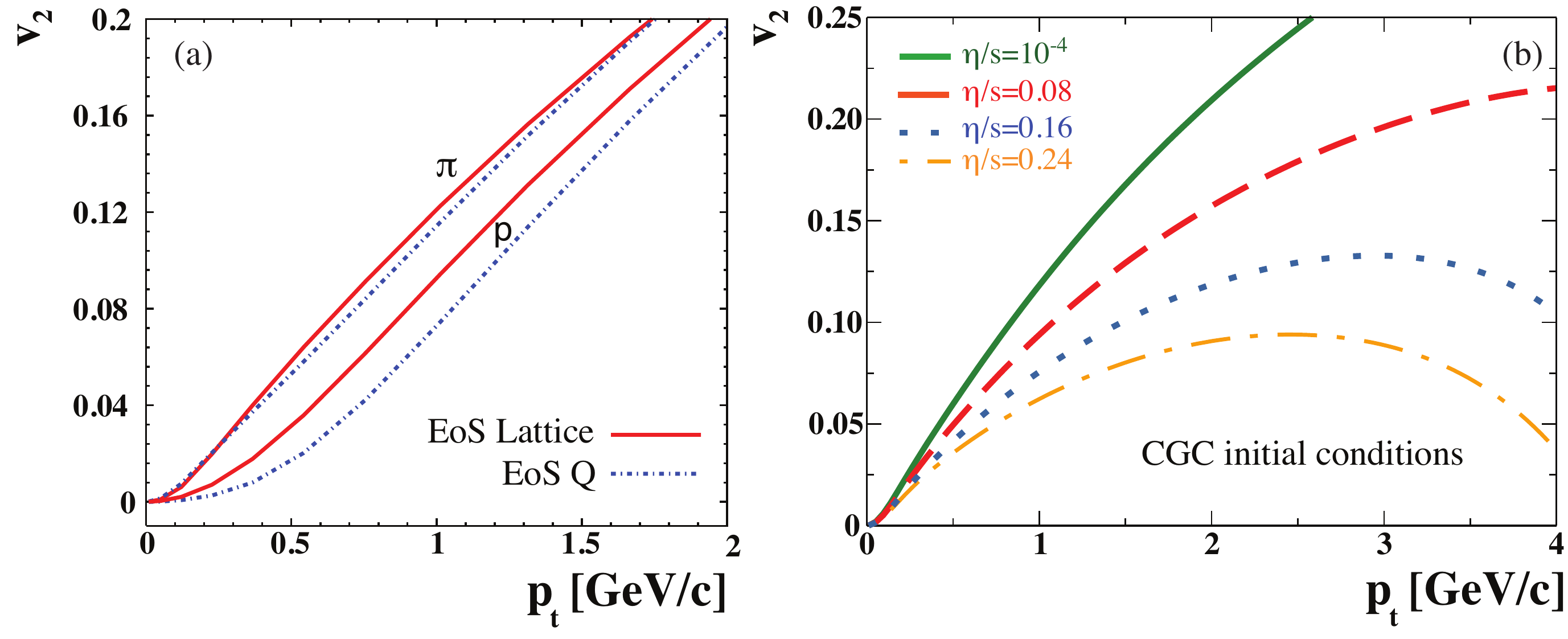}
    \caption{
    a) The {\sc EoS} dependence of $v_2(\pt)$ for pions and protons~\cite{Huovinen:2009yb}. 
    The full lines are for the lattice inspired EoS and the dashed lines for
    an EoS which incorporates a first order phase transition. 
    b) The dependence on $\eta$/s of $v_2(\pt)$ for charged particles~\cite{Luzum:2008cw}.
    }
    \label{v2pt_theory}
  }
\end{figure}
Although $\varepsilon_{p}$ is not a direct observable, 
the observed EoS dependence of $\varepsilon_{p}$ versus time is reflected in the experimental observable $v_2$, 
in particular when plotted as function of transverse momentum and particle mass.
Figure~\ref{v2pt_theory}a shows $p_{\rm t}$-differential elliptic flow for pions and protons after the transverse 
momentum spectra have been constrained. 
A clear mass dependence of $v_2$ at low transverse momentum 
is observed for both equations of state. The figure also clearly shows that the pion $v_2$ does not change much between the lattice EoS and EoS Q. 
On the other hand, the $v_2$ of protons does change significantly because the heavier particles are more sensitive to the change in collective
motion.
Therefore measurements of $v_{2}(p_{t})$ for various particle species provide an excellent constraint on the EoS in ideal hydrodynamics.

More recently, it was realized that small deviations from ideal hydrodynamics, in particular viscous corrections, already modify significantly the 
buildup of the elliptic flow~\cite{Teaney:2003kp}. 
The shear viscosity determines how good a fluid is\footnote{a good fluid has a small viscosity and does not convert much 
kinetic energy of the flow into heat.}, however,  for relativistic fluids the more useful quantity is the shear viscosity over entropy ratio $\eta/s$.
Known good fluids in nature have an $\eta/s$ of order $\hbar/k_{\rm B}$. 
In a strongly coupled ${\cal N} = 4$ supersymmetric Yang Mills theory with a large number of colors ('t Hooft limit), $\eta/s$ can be 
calculated using a gauge gravity duality~\cite{Kovtun:2004de}: 
\[
\frac{\eta}{s} = \frac{\hbar}{4\pi k_{\rm B}}.
\]
Kovtun, Son and Starinets conjectured, using the AdS/CFT correspondence, that this implies that 
all fluids have $\eta/s \geq \hbar/4\pi k_{\rm B}$ (the {\sc KSS} bound.). 
We therefore call a fluid with $\eta/s = 1/4\pi$ (in natural units) a perfect fluid. 
The {\sc KSS} bound raises the interesting question on how fundamental this value is in nature and 
if the {\sc QGP} behaves like an almost perfect fluid. 
It is argued that the transition from hadrons to quarks and gluons 
occurs in the vicinity of the minimum in $\eta/s$, just as is the case for the phase transitions in helium, nitrogen, and water.   
An experimental measurement of the minimal value of $\eta/s$ would thus pinpoint the location of the transition~\cite{Csernai:2006zz,Lacey:2006bc}.

Experimentally we might get an answer to the magnitude of $\eta/s$ by measuring $v_{2}$ as is shown in Fig.~\ref{v2pt_theory}b. 
The full line is close to ideal hydrodynamics ($\eta/s \sim 0$) while the three other lines correspond to $\eta/s$ values 
of up to three times the {\sc KSS} bound. 
Different magnitudes of $\eta/s$ clearly lead to a dramatically different magnitude of $v_{2}$ and change its $p_{\rm t}$ dependence.  
However the magnitude and $p_{\rm t}$ dependence of $v_{2}$ not only depend on $\eta/s$ but also on the EoS as we have seen. 

\begin{figure}[hbt]
  {\center
      \includegraphics[width=0.98\textwidth]{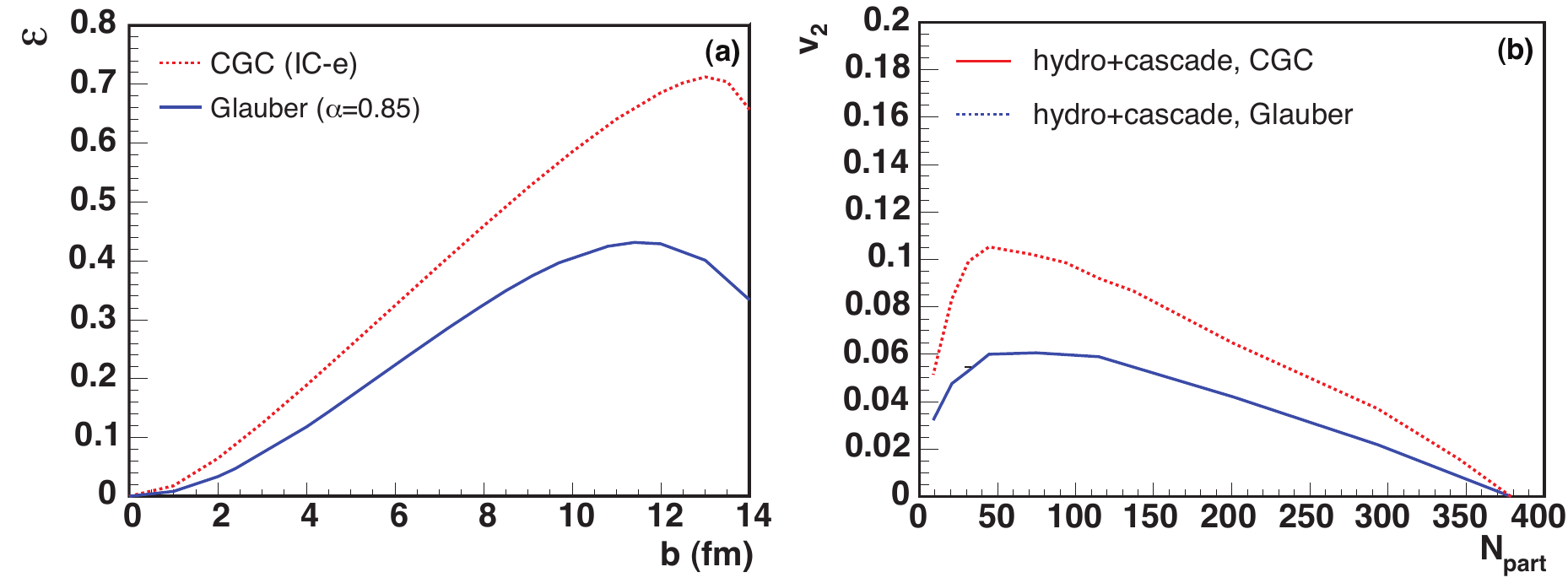}
    \caption{
    a) The eccentricity $\varepsilon$ calculated in a color glass condensate (CGC) model and using a Glauber model~\cite{Hirano:2007xd}.
    b) The $v_2$ obtained using the CGC or Glauber initial eccentricity~\cite{Hirano:2007xd}.
    }
    \label{v2_ecc}
  }
\end{figure}
The magnitude of $v_2$ does not only depend on the medium properties of interest, 
but is also proportional to the initial spatial anisotropy of the collision region. 
This spatial anisotropy can be characterized by the eccentricity, which is defined by
\begin{equation}
\varepsilon =  \frac{\left< y^{2} - x^{2}\right>}{\left<y^{2}+x^{2}\right>},
\label{eq:epsilon_std}
\end{equation} 
where $x$ and $y$ are the positions of the participating nucleons in the transverse plane and the brackets denote an average which 
traditionally was taken over the number of participants.  Recent calculations have shown that the eccentricity obtained in different descriptions, 
in particular comparing a Glauber with a Color Glass Condensate ({\sc CGC}) description, shows that $\varepsilon$ varies by almost 25\% 
at a given impact parameter~\cite{Hirano:2007xd}, see Fig.~\ref{v2_ecc}a. 
The elliptic flow, obtained when using these different initial eccentricities is shown in Fig.~\ref{v2_ecc}b. 
As expected, the different magnitude of the eccentricity propagates to the magnitude of the elliptic flow. 
Because currently we cannot measure the eccentricity independently 
this leads to a large uncertainty in experimental determination of $\eta$/s. 

To summarize,  we have seen that the elliptic flow depends on fundamental properties of the created matter, in particular the sound velocity and 
the shear viscosity, 
but also on the initial spatial eccentricity. 
Detailed measurements of elliptic flow as function of transverse momentum, particle mass and collision centrality  
provide an experimental handle on these properties. 
In the next section, before we discuss the measurements, 
we first explain how we estimate the anisotropic flow experimentally. 
 
\section{Elliptic Flow: Analysis Methods}
\label{methods}

Because the reaction plane angle is not a direct observable the elliptic flow (Eq.~\ref{fouriercoeff}) can not be measured directly
so that it is usually estimated using azimuthal correlations between the observed particles. 
Two-particle azimuthal correlations, for example, can be written as:
\begin{eqnarray}
\langle\langle e^{i2(\varphi_1 - \varphi_2)} \rangle\rangle &=& \langle\langle e^{i2(\varphi_1 - \Psi_{\rm RP} - (\varphi_2 - \Psi_{\rm RP}))} \rangle\rangle , 
\nonumber\\
&=& \langle\langle e^{i2(\varphi_1 - \Psi_{\rm RP})} \rangle \langle e^{-i2(\varphi_2 - \Psi_{\rm RP})} \rangle + \delta_{2}\rangle, 
\nonumber\\
&=& \langle v_2^2 + \delta_{2}\rangle, 
\label{twoParticleFlowEstimate3}
\end{eqnarray}
where the double brackets denote an average over all particles within an event, followed by averaging over all events.
In Eq.~\ref{twoParticleFlowEstimate3} we have factorized the azimuthal correlation 
between the particles in a common correlation with the reaction plane (elliptic flow $v_2$) and a correlation 
independent of the reaction plane (non-flow $\delta_2$). 
Here we have assumed that the correlation between $v_2$ and $\delta_2$ is negligible.
If $\delta_2$ is small, Eq.~\ref{twoParticleFlowEstimate3} can be used to measure $\left<v_{2}^{2}\right>$, but in general the non-flow
contribution is not negligible.
\begin{figure}[thb]
  {\center
      \includegraphics[width=0.9\textwidth]{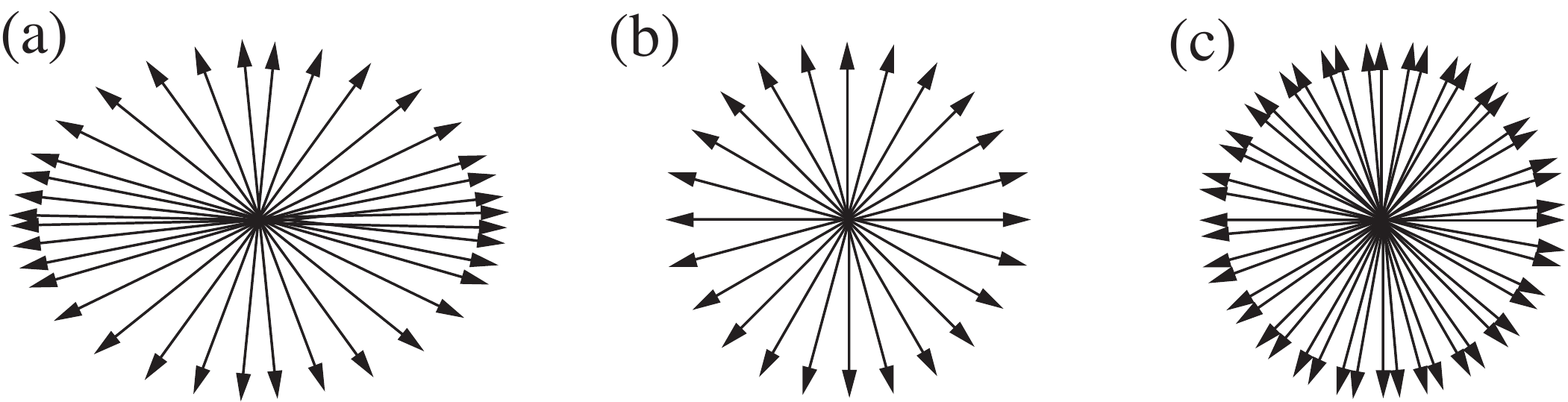}
    \caption{
    Examples of particle distributions in the transverse plane, where for 
    a) $v_{2} > 0$, $v_{2}\{2\} > 0$, 
    b) $v_{2} = 0$, $v_{2}\{2\} = 0$, and
    c) $v_{2} = 0$, $v_{2}\{2\} > 0$.
    }
    \label{2particle}
  }
\end{figure}
In Fig.~\ref{2particle} we illustrate two-particle nonflow contributions as follows: In Fig.~\ref{2particle}a an anisotropic distribution is shown
for which both $v_2 = \langle \cos 2 \phi \rangle$ and the two-particle correlation $v_2\{2\} = \sqrt{\langle \cos 2 (\phi_1 - \phi_2) \rangle}$ 
are positive. 
Figure~\ref{2particle}b shows a symmetric distribution for which $v_2 = 0$ and also $v_2\{2\} = 0$. 
Figure~\ref{2particle}c shows two symmetric distributions rotated with respect to each other which give $v_2 = 0$ while $v_2\{2\}$ is nonzero.
This illustrates how non-flow contributions from sources like resonance decays or jets can contribute to $v_2$ measured from two particle correlations. 
 
The collective nature of elliptic flow can be exploited to suppress non-flow contributions~\cite{Borghini:2001vi,Adler:2002pu}. 
This is done using so called cumulants, which are genuine multi-particle correlations. 
For instance, the two particle cumulant $c_{2}\{2\}$ and the four particle cumulants $c_{2}\{4\}$ are defined as:
\begin{eqnarray}
c_{2}\{2\} &\equiv& \left<\left<e^{i2(\varphi_1-\varphi_2)}\right>\right> = \left<v_2^2 + \delta_{2}\right>.
\label{twoParticleFlowEstimate4}\\
c_{2}\{4\} &\equiv& \left<\left<e^{i2(\varphi_{1} + \varphi_{2}-\varphi_{3}-\varphi_{4})}\right>\right>-2\left<\left<e^{i2(\varphi_1-\varphi_2)}\right>\right>^{2},
\nonumber\\
&=& \left<v_{2}^{4} + \delta_{4} + 4 v_{2}^{2}\delta_{2} + 2\delta_{2}^{2}\right> - 2\left< v_{2}^{2}+ \delta_{2}\right>^{2},  
\nonumber\\
&=& \left<-v_{2}^{4} + \delta_{4}\right>.
\label{fourParticleFlowEstimate3}
\end{eqnarray}
From the combinatorics it is easy to show that 
$\delta_{2} \propto 1/M_{\rm c}$ and $\delta_{4} \propto 1/M_{\rm c}^{3}$, where $M_{\rm c}$ is the number of independent particle clusters.
Therefore, $v_{2}\{2\}$ is only a good estimate if $v_{2} \gg 1/\sqrt{M_{\rm c}}$ while  $v_{2}\{4\}$ is already a good estimate of 
$v_{2}$ if $v_{2} \gg 1/{M_{\rm c}}^{3/4}$; for $c_2\{\infty\}$ this argument leads to $v_{2} \gg 1/M_{\rm c}$. 
This shows that for a typical Pb--Pb collision at the LHC with $M_{\rm c} = 500$ the possible non-flow contribution can be reduced by more than an order of magnitude using higher order cumulants.  
One of the problems in using multi-particle correlations is the computing power needed to go over all possible particle multiplets. 
To avoid this problem, multi-particle correlations in heavy-ion collision are calculated from generating functions with 
numerical interpolations~\cite{Borghini:2001vi} or, as was shown more recently, from an exact solution~\cite{Bilandzic:2010jr}. 

The last equality in Eq.~\ref{fourParticleFlowEstimate3} follows from the assumption that $v_2$ and $\delta_2$ are uncorrelated and also 
that $\langle \delta_{2}^{2} \rangle = \langle \delta_{2} \rangle^2$ and 
$\langle v_{2}^{4} \rangle = \langle v_{2}^{2} \rangle^2$.
In other words, we have neglected the event-by-event fluctuations in $v_2$ and $\delta_2$.
The effect of the fluctuations on $v_2$ estimates can be obtained from
\begin{eqnarray}
    \langle v_2^2 \rangle &=& \langle v_2 \rangle^2 + \sigma^2, \nonumber \\
    \langle v_2^4 \rangle &=& \langle v_2 \rangle^4 + 6\sigma^2 \langle v_2 \rangle^2, \nonumber \\
    \langle v_2^6 \rangle &=& \langle v_2 \rangle^6 + 15\sigma^2 \langle v_2 \rangle^4. \label{sigma} 
\end{eqnarray}
Neglecting the non-flow terms we have the following expressions for the cumulants:
\begin{eqnarray}
    v_2\{2\} &=& \sqrt{\langle v_2^2 \rangle}, \nonumber \\
    v_2\{4\} &=& \sqrt[4]{2 \langle v_2^2\rangle^2 - \langle v_2^4 \rangle}, \nonumber \\
    v_2\{6\} &=& \sqrt[6]{\frac{1}{4} \left( \langle v_2^6 \rangle -9 \langle v_2^2\rangle \langle v_2^4\rangle + 12 \langle v_2^2\rangle^3 \right)}. 
    \label{fcummu}
\end{eqnarray}
Here we have introduced the notation $v_2\{n\}$ as the flow estimate from the cumulant $c_2\{n\}$.
Assuming that $\sigma \ll \langle v\rangle$ we obtain from Eqs.~\ref{sigma} and~\ref{fcummu}, up to order $\sigma^2$:
\begin{eqnarray}
    v_2\{2\} &=& \langle v_2 \rangle + \frac{1}{2} \frac{\sigma^2}{\langle v_2 \rangle},\nonumber \\
    v_2\{4\} &=& \langle v_2 \rangle - \frac{1}{2} \frac{\sigma^2}{\langle v_2 \rangle},\nonumber \\ 
    v_2\{6\} &=&  \langle v_2 \rangle - \frac{1}{2} \frac{\sigma^2}{\langle v_2 \rangle}.
    \label{fluctuations}
\end{eqnarray}
From Eqs.~\ref{twoParticleFlowEstimate4} and~\ref{fluctuations} it is clear that the difference between $v_{2}\{2\}$ and $v_{2}\{4\}$ is 
sensitive to non-flow and fluctuations.

\begin{figure}[thb]
  {\center
      \includegraphics[width=0.5\textwidth]{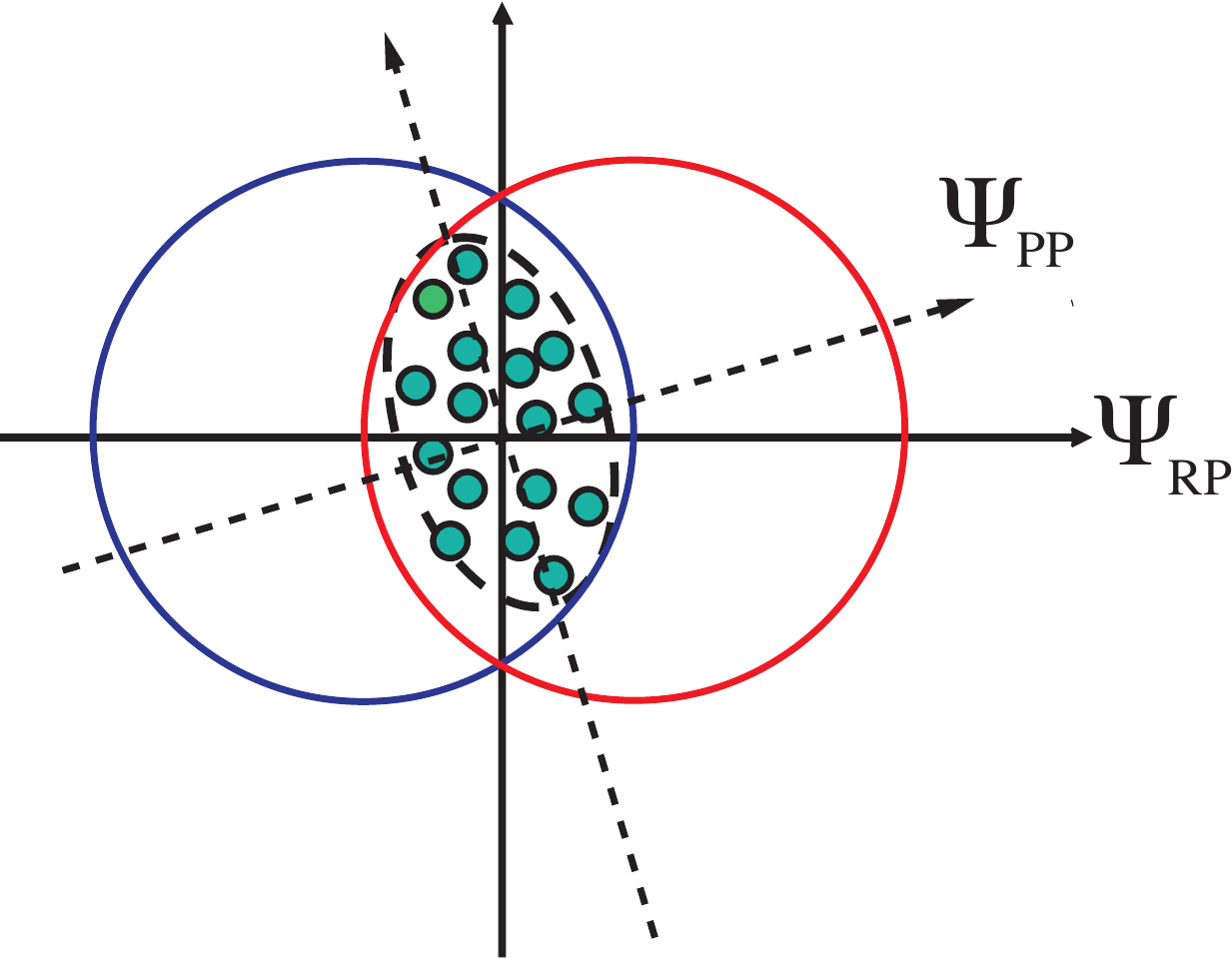}
    \caption{
 Transverse view of a heavy-ion collision with the reaction plane $\Psi_{\rm RP}$ oriented along the $x$-axis.
 Indicated are the participants in the overlap region that randomly define a particpant plane $\Psi_{\rm PP}$ for each collision.
     }
    \label{planes}
  }
\end{figure}
Flow fluctuations have become an important part of elliptic flow 
studies~\cite{Miller:2003kd,Manly:2005zy,Sorensen:2009cz,Qin:2010pf,Lacey:2010hw,Hirano:2009bd,Ollitrault:2009ie,Mrowczynski:2009wk,
Bhalerao:2006tp,Voloshin:2007pc,Alver:2008zza}. 
It is believed that such fluctuations originate mostly from fluctuations in the 
initial collision geometry.
This is illustrated in Fig.~\ref{planes} which shows participants that are randomly distributed in the overlap region.
This collection of participants defines a participant plane $\Psi_{\rm PP}$~\cite{Manly:2005zy}  which fluctuates, 
for each event, around the reaction plane $\Psi_{\rm RP}$.
These fluctuations can be estimated from calculations in, for instance,  a Glauber model. 

\begin{figure}[thb]
  {\center
      \includegraphics[width=0.98\textwidth]{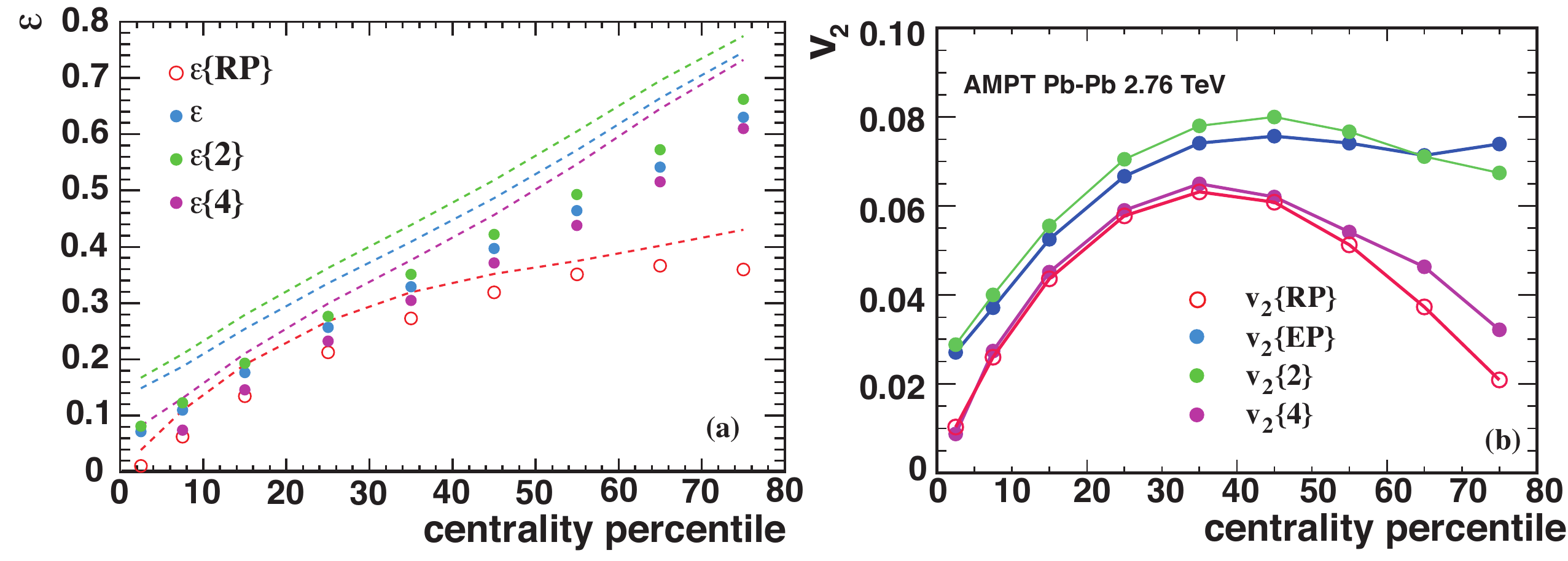}
    \caption{
    a) The eccentricities from a Glauber calculation for participating nucleons (the solid and open markers) and binary
    collisions (the dashed lines).
    b) Various $v_2$ estimates  compared to the reaction plane value, $v_2\{{\rm RP}\}$.
    }
    \label{eccentricities}
  }
\end{figure}
Figure~\ref{eccentricities}a shows the eccentricities (Eq.~\ref{eq:epsilon_std}) calculated in a Glauber model. 
Here $\varepsilon\{{\rm RP}\}$ denotes the eccentricity in the reaction plane, 
$\varepsilon$ is the participant eccentricity and $\varepsilon\{2\}$ and $\varepsilon\{4\}$ are the participant eccentricities calculated 
using the cumulants, analogous to the definitions in Eq.~\ref{fcummu}~\cite{Miller:2003kd}.
In Fig.~\ref{eccentricities}a the eccentricities are calculated using as a weight the participating nucleons (open and solid markers) or  
as a weight binary collisions (dashed lines). 
The figure clearly shows that in both cases $\varepsilon$ is in between $\varepsilon\{2\}$ and $\varepsilon\{4\}$ as is expected 
from Eq.~\ref{fluctuations}. The figure also shows that $\varepsilon\{4\}$ is close to $\varepsilon\{{\rm RP}\}$ for the $0-40$\% 
centrality range~\cite{Bhalerao:2006tp,Voloshin:2007pc}.
In Fig.~\ref{eccentricities}b we show a transport model calculation of $v_2$ in the {\sc AMPT} model~\cite{Lin:2004en}. 
In this model the true reaction plane is known so that we can compare the different $v_2$ estimates with the value in the 
reaction plane. The {\sc AMPT} model uses a Glauber model for the initial conditions and we can therefore compare these estimates 
with Fig.~\ref{eccentricities}a (the dashed lines). 
The agreement between $v_{2}\{4\}$ and $v_{2}\{{\rm RP}\}$ holds for most of the centrality range, 
while for the eccentricities in the Glauber model a large difference is observed for the more peripheral collisions~\cite{Alver:2008zza}. 

\section{Elliptic Flow Measurements}
\label{measurements}

\subsection{The Perfect Liquid}

The large elliptic flow observed at RHIC provides compelling evidence for strongly interacting matter
which appears to behave like an almost perfect liquid~\cite{experiments,Gyulassy:2004zy}. 
To quantify the agreement with an almost perfect fluid the significant viscous corrections need to be calculated. 
Based on different model assumptions the ratio  $\eta/s$ has been estimated at  \snn~=~200 GeV and  
is found to be below five times the {\sc KSS} bound~\cite{Teaney:2009qa,Shen:2010uy,Song:2010mg,Masui:2009pw,Nagle:2009ip}. 

\begin{figure}[thb]
  {\center
      \includegraphics[width=0.98\textwidth]{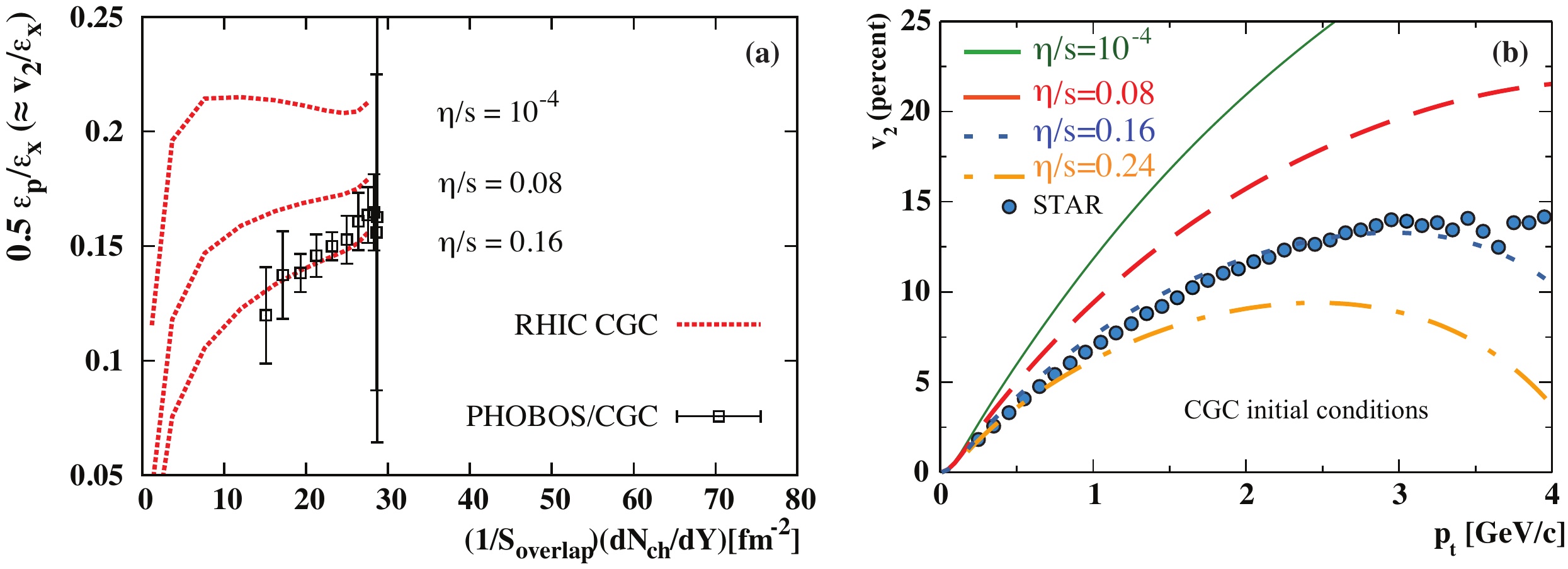}
    \caption{
    a) The centrality dependence of $v_2\{2\}$ compared to viscous hydrodynamic model calculations~\cite{Luzum:2008cw}.
    b)  The transverse momentum dependence of $v_2$ compared to  the same viscous hydrodynamic calculations~\cite{Luzum:2008cw}.
    }
    \label{viscous_hydro}
  }
\end{figure}
In Fig.~\ref{viscous_hydro} we show the centrality and transverse momentum dependence of $v_2$  compared to 
viscous hydrodynamic calculations~\cite{Luzum:2008cw} with different values of $\eta/s$. 
Using an eccentricity from a {\sc CGC} inspired calculation, it is seen 
that both the centrality and transverse momentum dependence is well described with an $\eta/s$ of 
two times the {\sc KSS} bound. These calculations are performed under the  assumption that  
the value of $\eta/s$ is constant during the entire evolution. 
The value used in these calculations should be considered as an effective average of $\eta/s$, 
because we know from other fluids that $\eta/s$ depends on temperature.  
In addition, we also know that part of the elliptic flow originates from the hadronic phase. 
Therefore, knowledge of the temperature dependence {\it and} knowledge of the relative contributions 
from the partonic and hadronic phase is required to quantify $\eta/s$ of the partonic fluid. 

\begin{figure}[thb]
  {\center
      \includegraphics[width=0.98\textwidth]{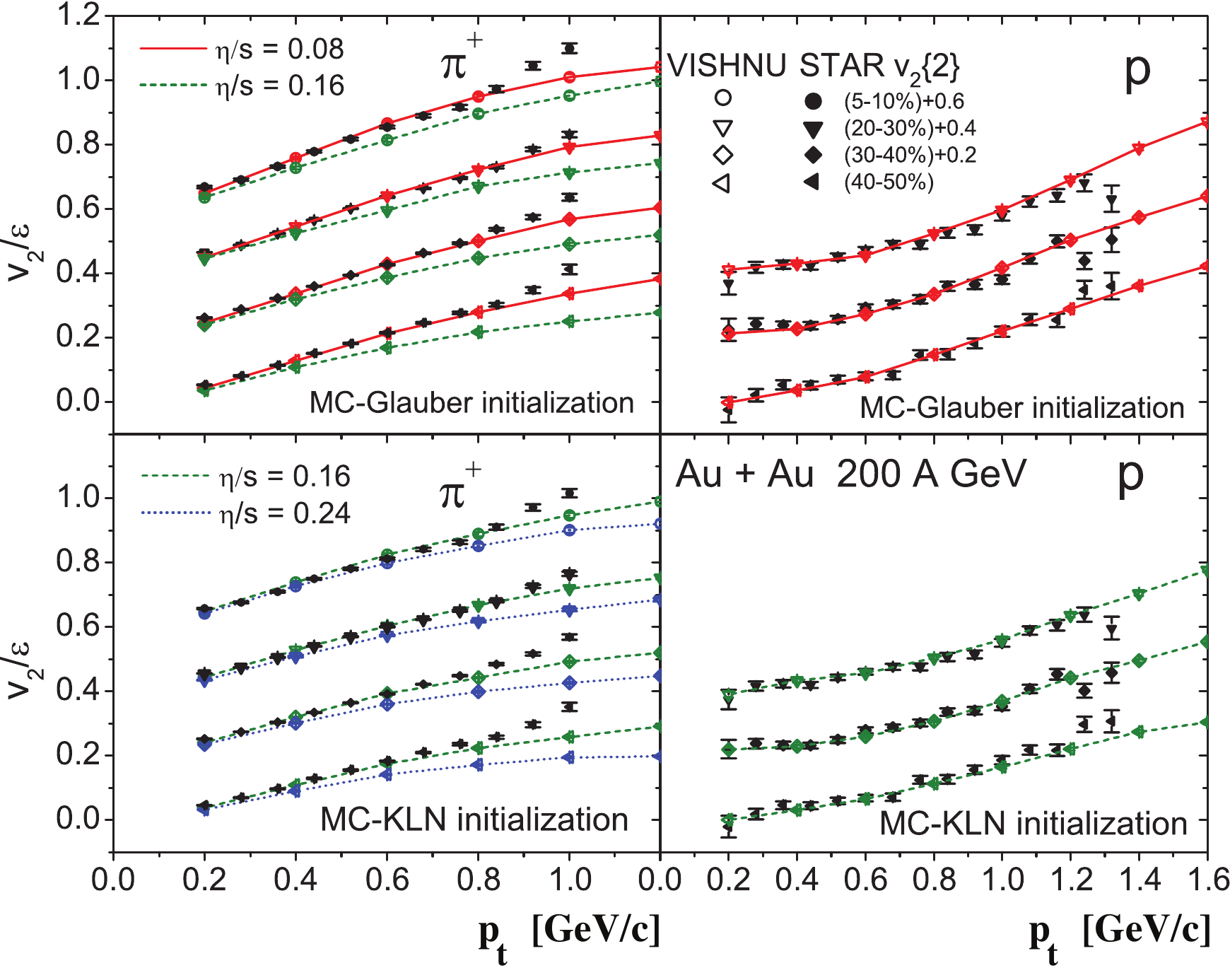}
    \caption{
    The $v_2(\pt)$ for pions and protons measured by STAR compared to model calculations 
    with different eccentricities and $\eta/s$~\cite{Song:2011hk}.
    }
    \label{pid_v2pt}
  }
\end{figure}
Not only the $v_2$ of charged particles but also that of identified particles at RHIC is described in the framework of 
viscous hydrodynamics at low-\pt. 
Figure~\ref{pid_v2pt} shows the measured pion and proton elliptic flow measured by STAR compared to  {\sc VISHNU}~\cite{Song:2011hk} 
model calculations.
The {\sc VISHNU} model is a hybrid model which uses viscous hydrodynamics for the initial stage followed by a hadron 
cascade afterburner. In the initial viscous hydrodynamic stage $\eta/s$ is temperature independent. 
The $\eta/s$ magnitude required to describe the pion and proton elliptic flow data are found to be one or two times the {\sc KSS} bound 
for a Glauber or {\sc CGC} eccentricity, respectively. 
This is in agreement with the magnitude of $\eta/s$ required to describe charged particle $v_2$.

While the description of $v_2$ measurements  is encouraging, it is important to realize that 
there are still large uncertainties in: 
(i) the initial eccentricity, 
(ii) the relative contributions from the hadronic and partonic phase
and (iii) the temperature dependence of $\eta/s$. 
Elliptic flow measurements at the {\sc LHC}, with a higher center of mass energy, will constrain 
these uncertainties and will eventually provide a decisive test which of the currently successful model descriptions is the more appropriate.

\subsection{Energy Dependence}

Lead-Lead collisions at the LHC are expected to produce a system which is hotter and has a longer lived partonic phase
than the system created in Au-Au collisions at RHIC energies. 
As a consequence, the hadronic contribution to the elliptic flow decreases which reduces the 
uncertainty on the determination of $\eta/s$ in the partonic fluid. 
Because $\eta/s$ is expected to depend on temperature in both the partonic and hadronic system it was not clear if the 
elliptic flow would increase or decrease in going from RHIC to LHC energies. 
Hydrodynamic models~\cite{Niemi:2008ta,Kestin:2008bh,Luzum:2009sb} and 
hybrid models~\cite{Hirano:2010jg,Hirano:2005xf} that successfully describe flow at RHIC 
predicted an increase of $\sim$10--30\% in $v_2$. 

\begin{figure}[thb]
  {\center
      \includegraphics[width=0.98\textwidth]{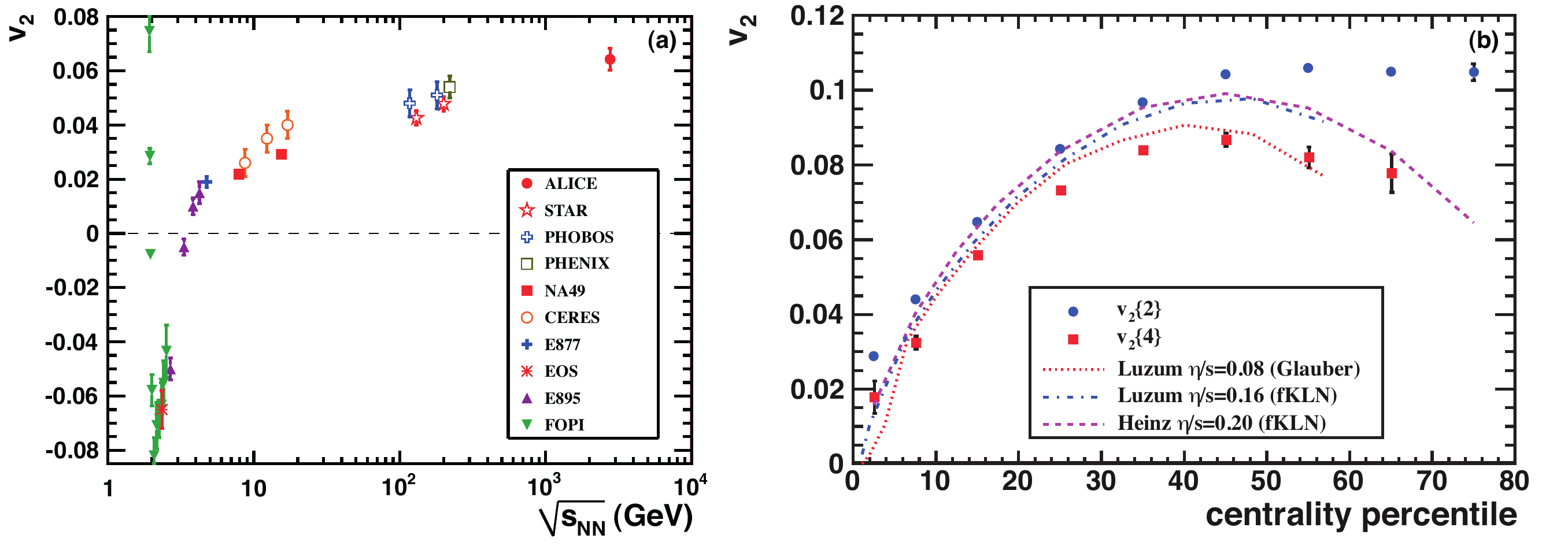}
    \caption{
    a) Integrated elliptic flow at 2.76~TeV in the 20--30\% centrality class 
    compared with results from lower energies taken at similar centralities (From~\cite{Aamodt:2010pa}). 
    b)  Elliptic flow as a function of event centrality, 
    for the 2- (full circles ) and 4-particle (full squares) cumulant methods compared to viscous hydrodynamic calculations
    (dashed lines)~\cite{Aamodt:2010pa,Luzum:2010ag,Shen_priv}. 
    }
    \label{v2edep_summary}
  }
\end{figure}
Figure~\ref{v2edep_summary}a shows the measured integrated elliptic flow at the LHC in one centrality bin, 
compared to results from lower energies. 
It shows that there is a continuous increase in the elliptic flow 
from {\sc RHIC} to {\sc LHC} energies.  In comparison to the elliptic flow measurements in Au--Au collisions at
$\snn~=~200$~GeV $v_2$ increases by about 30\% at $\snn~=~2.76$~TeV. 

Figure~\ref{v2edep_summary}b shows the $v_{2}$ for different centralities measured by ALICE 
with the two- and four-particle cumulant method. 
The difference between the two- and four-particle flow estimates for the more central collisions ($< 40$\%) is expected to be dominated by 
event-by-event flow fluctuations (see Eq.~\ref{fluctuations}). For the more peripheral collisions the two-particle cumulant is likely biased by non-flow.
We already mentioned that $v_2\{4\}$ yields estimates of the elliptic flow in the reaction plane which can thus be compared to model 
predictions of $v_2\{{\rm RP}\}$. 
The curves in Fig.~\ref{v2edep_summary}b show $v_2\{{\rm RP}\}$ from hydrodynamic model calculations for $\snn~=~2.76$~TeV, 
with initial eccentricities and magnitudes of $\eta/s$ which described the {\sc RHIC} data. 
It is seen that in hydrodynamic calculations the observed increase in $v_2$ from {\sc RHIC} to {\sc LHC} energies is 
within expectations. 
Detailed comparisons, however, have to wait till measurements of identified particle spectra and identified particle elliptic flow 
become available. 
It will then be important to see if one still obtains a quantitative description of the data in viscous hydrodynamics and 
what the required magnitude of $\eta/s$ then will be.

\begin{figure}[thb]
  {\center
      \includegraphics[width=0.98\textwidth]{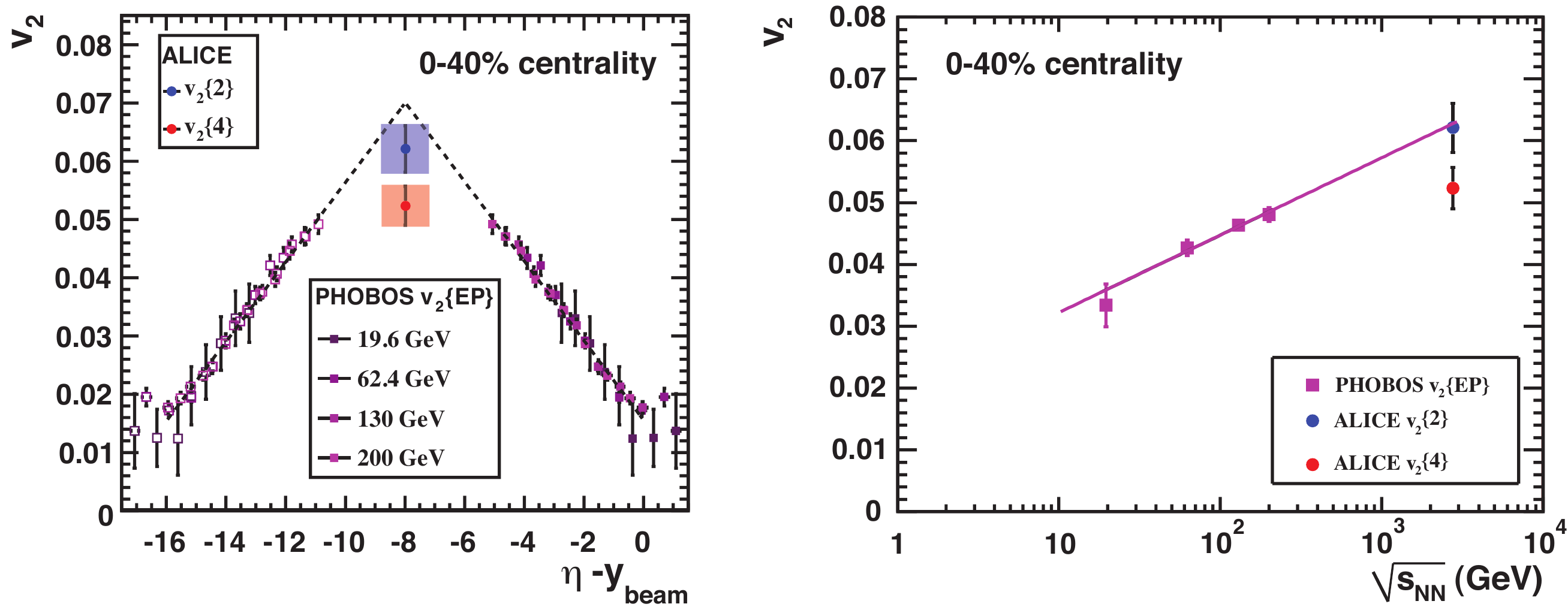}
    \caption{
    a) Elliptic flow plotted versus  $\eta - y_{{\rm beam}}$ for {\sc RHIC} and the {\sc LHC}, 
    with for $y_{{\rm beam}}$ the $\snn~=~2.76$~TeV value. The mesh shows the $\eta$-range over which
    the elliptic flow is constant within statistical uncertainties.
    b) Elliptic flow versus beam energy. 
    In both figures the uncertainties for {\sc ALICE} are systematic uncertainties and  
    the $v_2$ in 0--40\% centrality is obtained by averaging over events instead of over, the more commonly used, particle yield.
    }
    \label{simple_scaling}
  }
\end{figure}
In addition to comparisons with detailed dynamic model calculations we might also learn something from 
what happens to the several simple scaling properties observed at lower energies. 
For instance, it was shown that the integrated elliptic flow depends linearly on the pseudorapidity $\eta$, 
measured with respect to the beam rapidity $y_{\rm beam}$~\cite{Back:2004zg}, as is shown in Fig.~\ref{simple_scaling}a. 
Based on this scaling behaviour a phenomenological extrapolation~\cite{Busza:2007ke} from {\sc RHIC} 
to the {\sc LHC} was made (dashed line in Fig.~\ref{simple_scaling}a) which predict an increase in $v_2$ of $\sim$50\%, 
larger than predictions from most other models. 
{\sc PHOBOS} measured $v_2$ down to $p_t = 0$ using the eventplane method which at {\sc RHIC} is 
similar to measuring $v_2\{2\}$. 
The measurements at $\eta=0$ for the LHC are below the triangular extrapolation. 
However, the elliptic flow as function of pseudorapidity measured by 
ALICE in $|\eta| < 0.8$ (mesh in Fig.~\ref{simple_scaling}a) is constant within uncertainties. 
If one takes into account that the $v_2(\eta)$ does saturate, like the multiplicity, at each energy around midrapidity then the 
longitudinal scaling might hold up to {\sc LHC} energies. 

Figure~\ref{simple_scaling}b shows that the energy dependence of elliptic flow at midrapidity 
also seems to follow a rather simple scaling: the measured elliptic flow for four beam energies at {\sc RHIC} shows a linear increase, 
extrapolating this to $\snn~=~2.76$~TeV results in a $v_2$ which 
agrees well with the {\sc ALICE} measurement.
 
\begin{figure}[thb]
  {\center
      \includegraphics[width=0.98\textwidth]{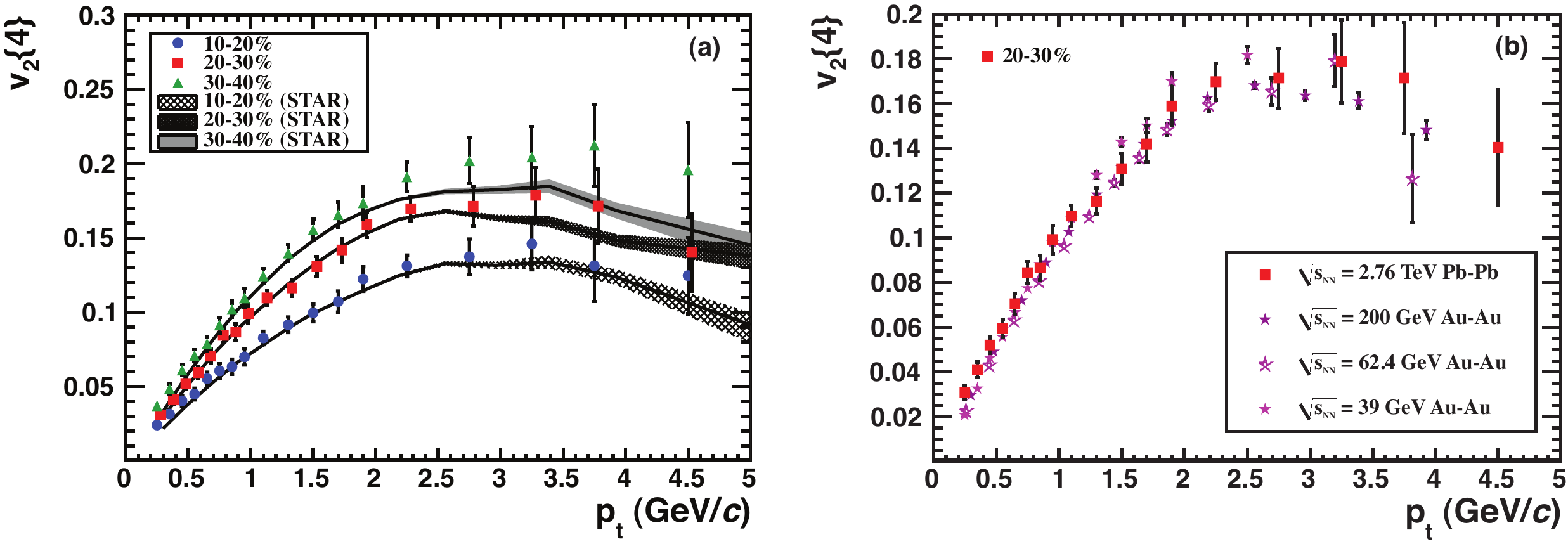}
    \caption{
    a) $v_2$\{4\}(\pt) for various centralities compared to STAR
    measurements. The data points in the 20--30\% centrality bin 
    are shifted in $p_{\rm t}$ for visibility.
    b)  $v_2$\{4\}(\pt) for a collision energy range which covers almost two orders of magnitude~\cite{Kumar:2011de} 
    }
    \label{compare_v2pt}
  }
\end{figure}
The observed increase in the elliptic flow as function of beam energy is either due to an increase in $\pt$-differential flow 
or due to an increase in the average transverse momentum of the charged particles.
In most hydrodynamic model calculations the $\pt$-differential elliptic flow of charged particles does not change 
significantly~\cite{Niemi:2008ta,Kestin:2008bh}, 
while the radial (azimuthally symmetric) flow does increase which leads to an increase in the average transverse momentum. 
The larger radial flow also leads to a decrease of the elliptic flow at low transverse momentum, which is most pronounced for heavy particles. 
Figure~\ref{compare_v2pt}a compares the $\pt$-differential elliptic flow of charged particles for three centralities at the {\sc LHC} with {\sc STAR} 
measurements at {\sc RHIC}. It is seen that the $\pt$-differential elliptic flow is the same within experimental uncertainties. 
In Fig.~\ref{compare_v2pt}b the $\pt$-differential elliptic flow measured at four beam energies is shown.
The agreement of $v_2(\pt)$ at these beam energies, which differ by almost two orders of magnitude, is remarkable.
Measurements of identified particle elliptic flow at these energies will reveal if this agreement can be understood in hydrodynamic model calculations.
 
\section{Summary}
In this review, I have shown that elliptic flow is one of the most informative observables in heavy-ion collisions.
Nevertheless, the wealth of experimental information obtained from elliptic flow is far from being fully explored. 
The theoretical understanding of the experimental data is rapidly improving as is our understanding 
of the dynamics in heavy-ion collisions and the
properties of the new state of matter, the quark gluon plasma. 
New high quality data from the {\sc LHC} recently became available which shows that at {\sc LHC} energies 
elliptic flow can be studied with unprecedented precision. 
This is because of the increase in particle multiplicity and also because of the increase in the flow signal itself. 
Due to  the expected longer life time of the quark gluon plasma and the smaller contributions from the hadronic phase 
it is argued that the {\sc LHC} is also better suited to determine $\eta/s$ of the partonic fluid~\cite{Niemi:2011ix}. 
Measurements of identified particle elliptic flow at the {\sc LHC} and in particular the stronger mass dependence 
(splitting) of $v_2(\pt)$ will be important to confirm the current theoretical picture.
Additional constraints on $\eta/s$ can be obtained by measurements of the other anisotropic flow harmonics $v_3$, $v_4$ and $v_5$.
In the near future these measurements will become available and significantly increase our understanding of  ultra-relativistic nuclear
collisions and multi-particle production in general.

\section*{Acknowledgements}

The author would like to thank Ante~Bilandzic, Michiel~Botje, Pasi~Huovinen and You~Zhou for their contributions.
This work is supported by NWO and FOM.
\\


\begin{thebibliography}{99}

\bibitem{Higgs:1964ia}
Peter~W. Higgs.
\newblock Phys. Lett. 12:132--133, 1964.

\bibitem{Higgs:1964pj}
Peter~W. Higgs.
\newblock Phys. Rev. Lett. 13:508--509, 1964.

\bibitem{Englert:1964et}
F.~Englert and R.~Brout.
\newblock Phys. Rev. Lett. 13:321--322, 1964.

\bibitem{Nambu:1961tp}
Yoichiro Nambu and G.~Jona-Lasinio.
\newblock Phys. Rev. 122:345--358, 1961.

\bibitem{Gross:1973ju}
D.~J. Gross and Frank Wilczek.
\newblock Phys. Rev. D8:3633--3652, 1973.

\bibitem{Politzer:1973fx}
H.~David Politzer.
\newblock Phys. Rev. Lett. 30:1346--1349, 1973.


\bibitem{Shuryak:1980tp}
Edward~V. Shuryak.
\newblock Phys. Rept. 61:71--158, 1980.

\bibitem{Bazavov:2009zn}
A.~Bazavov, T.~Bhattacharya, M.~Cheng {\it et al.},
Phys.\ Rev.\  {\bf D80}, 014504 (2009).

\bibitem{Chapline:1974zf}
G.~F. Chapline, M.~H. Johnson, E.~Teller, and M.~S. Weiss.
\newblock Phys. Rev. D8:4302--4308, 1973.

\bibitem{Lee:1974ma}
T.~D. Lee and G.~C. Wick.
\newblock Phys. Rev. D9:2291, 1974.

\bibitem{Lee:1978mf}
T.~D. Lee.
\newblock Phys. Rev. D19:1802, 1979.

\bibitem{Collins:1974ky}
John~C. Collins and M.~J. Perry.
\newblock Phys. Rev. Lett. 34:1353, 1975.

\bibitem{Pisarski:1983ms}
Robert~D. Pisarski and Frank Wilczek.
\newblock Phys. Rev. D29:338--341, 1984.

\bibitem{Ollitrault:1992bk}
 J.~Y.~Ollitrault,
 Phys.\ Rev.\  D {\bf 46}, 229 (1992).

\bibitem{Voloshin:2008dg}
  S.~A.~Voloshin, A.~M.~Poskanzer and R.~Snellings,
  in Landolt-Boernstein, {\em Relativistic Heavy Ion Physics}, Vol. 1/23
  (Springer-Verlag, 2010), p 5-54.
arXiv:0809.2949 [nucl-ex].  
  
\bibitem{Heinz:2009xj}
  U.~W.~Heinz,
  arXiv:0901.4355 [nucl-th].

\bibitem{Huovinen:2006jp}
  P.~Huovinen and P.~V.~Ruuskanen,
  Ann.\ Rev.\ Nucl.\ Part.\ Sci.\  {\bf 56}, 163 (2006)
   
\bibitem{Teaney:2009qa}
  D.~A.~Teaney,
  arXiv:0905.2433 [nucl-th].
     
\bibitem{Miller:2007ri}
  M.~L.~Miller, K.~Reygers, S.~J.~Sanders {\it et al.},
  Ann.\ Rev.\ Nucl.\ Part.\ Sci.\  {\bf 57}, 205-243 (2007).

\bibitem{Aamodt:2010pa}
  K.~Aamodt {\it et al.}  [The ALICE Collaboration],
  Phys.\ Rev.\ Lett.\  {\bf 105}, 252302 (2010).

\bibitem{Kolb:2003dz}
  P.~F.~Kolb, U.~W.~Heinz,
  In *Hwa, R.C. (ed.) et al.: Quark gluon plasma* 634-714.

\bibitem{Huovinen:2009yb}
  P.~Huovinen, P.~Petreczky,
  Nucl.\ Phys.\  {\bf A837}, 26-53 (2010).

\bibitem{Teaney:2003kp}
  D.~Teaney,
  Phys.\ Rev.\  {\bf C68}, 034913 (2003).
   
\bibitem{Luzum:2008cw}
  M.~Luzum, P.~Romatschke,
  Phys.\ Rev.\  {\bf C78}, 034915 (2008).
         
\bibitem{Kovtun:2004de}
  P.~Kovtun, D.~T.~Son, A.~O.~Starinets,
  Phys.\ Rev.\ Lett.\  {\bf 94}, 111601 (2005).

\bibitem{Csernai:2006zz}
  L.~P.~Csernai, J.~.I.~Kapusta, L.~D.~McLerran,
  Phys.\ Rev.\ Lett.\  {\bf 97}, 152303 (2006).

\bibitem{Lacey:2006bc}
  R.~A.~Lacey, N.~N.~Ajitanand, J.~M.~Alexander {\it et al.},
  Phys.\ Rev.\ Lett.\  {\bf 98}, 092301 (2007).

\bibitem{Hirano:2007xd}
  T.~Hirano, U.~W.~Heinz, D.~Kharzeev {\it et al.},
  J.\ Phys.\ G {\bf G34}, S879-882 (2007).

\bibitem{Borghini:2001vi}
  N.~Borghini, P.~M.~Dinh, J.~-Y.~Ollitrault,
  Phys.\ Rev.\  {\bf C64}, 054901 (2001).

\bibitem{Adler:2002pu}
  C.~Adler {\it et al.} [ STAR Collaboration ],
  Phys.\ Rev.\  {\bf C66}, 034904 (2002).
  
\bibitem{Bilandzic:2010jr}
  A.~Bilandzic, R.~Snellings and S.~Voloshin,
  arXiv:1010.0233 [nucl-ex], submitted to PRC

\bibitem{Miller:2003kd}
 M.~Miller and R.~Snellings,
 arXiv:nucl-ex/0312008.

\bibitem{Manly:2005zy}
 S.~Manly {\it et al.}  [PHOBOS Collaboration],
 Nucl.\ Phys.\  A {\bf 774}, 523 (2006).

\bibitem{Sorensen:2009cz}
  P.~Sorensen,
  [arXiv:0905.0174 [nucl-ex]].
  
\bibitem{Qin:2010pf}
  G.~-Y.~Qin, H.~Petersen, S.~A.~Bass {\it et al.},
  Phys.\ Rev.\  {\bf C82}, 064903 (2010).
 
 \bibitem{Lacey:2010hw}
  R.~A.~Lacey, R.~Wei, N.~N.~Ajitanand {\it et al.},
  [arXiv:1009.5230 [nucl-ex]].

\bibitem{Hirano:2009bd}
  T.~Hirano, Y.~Nara,
  Nucl.\ Phys.\  {\bf A830}, 191C-194C (2009).
  
\bibitem{Ollitrault:2009ie}
  J.~-Y.~Ollitrault, A.~M.~Poskanzer, S.~A.~Voloshin,
  Phys.\ Rev.\  {\bf C80}, 014904 (2009).
  
  \bibitem{Mrowczynski:2009wk}
  S.~Mrowczynski,
  Acta Phys.\ Polon.\  {\bf B40}, 1053-1074 (2009).

\bibitem{Bhalerao:2006tp}
  R.~S.~Bhalerao, J.~-Y.~Ollitrault,
  Phys.\ Lett.\  {\bf B641}, 260-264 (2006).
   
\bibitem{Voloshin:2007pc}
  S.~A.~Voloshin, A.~M.~Poskanzer, A.~Tang {\it et al.},
  Phys.\ Lett.\  {\bf B659}, 537-541 (2008).

\bibitem{Alver:2008zza}
  B.~Alver, B.~B.~Back, M.~D.~Baker {\it et al.},
  Phys.\ Rev.\  {\bf C77}, 014906 (2008).

\bibitem{Lin:2004en}
  Z.~-W.~Lin, C.~M.~Ko, B.~-A.~Li {\it et al.},
  Phys.\ Rev.\  {\bf C72}, 064901 (2005).
 
 \bibitem{experiments}
  I.~Arsene {\it et al.},
  Nucl.\ Phys.\  {\bf A757}, 1 (2005);
  B.~B.~Back {\it et al.},
  {\it ibid.,} p.\,28;
  J.~Adams {\it et al.},
  {\it ibid.,} p.\,102;
  K.~Adcox {\it et al.},
  {\it ibid.,} p.\,184.

\bibitem{Gyulassy:2004zy}
M.~Gyulassy, L.~McLerran,
Nucl.\ Phys.\  A {\bf 750}, 30 (2005).

\bibitem{Shen:2010uy}
  C.~Shen, U.~Heinz, P.~Huovinen {\it et al.},
  Phys.\ Rev.\  {\bf C82}, 054904 (2010).
 
 \bibitem{Song:2010mg}
  H.~Song, S.~A.~Bass, U.~W.~Heinz {\it et al.},
  [arXiv:1011.2783 [nucl-th]].
 
\bibitem{Masui:2009pw}
  H.~Masui, J-Y.~Ollitrault, R.~Snellings {\it et al.},
  Nucl.\ Phys.\  {\bf A830}, 463C-466C (2009).

\bibitem{Nagle:2009ip}
  J.~L.~Nagle, P.~Steinberg, W.~A.~Zajc,
  Phys.\ Rev.\  {\bf C81}, 024901 (2010).
 
\bibitem{Song:2011hk}
  H.~Song, S.A~Bass, U.W~Heinz {\it et al.},
  [arXiv:1101.4638 [nucl-th]].
              
\bibitem{Niemi:2008ta}
  H.~Niemi, K.~J.~Eskola and P.~V.~Ruuskanen,
  Phys.\ Rev.\  C {\bf 79}, 024903 (2009)
 
\bibitem{Kestin:2008bh}
  G.~Kestin and U.~W.~Heinz,
  Eur.\ Phys.\ J.\  C {\bf 61}, 545 (2009)

\bibitem{Luzum:2009sb}
  M.~Luzum and P.~Romatschke,
  Phys.\ Rev.\ Lett.\  {\bf 103}, 262302 (2009)
  
\bibitem{Hirano:2010jg}
  T.~Hirano, P.~Huovinen and Y.~Nara,
  arXiv:1010.6222 [nucl-th].

\bibitem{Hirano:2005xf}
  T.~Hirano, U.~W.~Heinz, D.~Kharzeev, R.~Lacey and Y.~Nara,
  Phys.\ Lett.\  B {\bf 636}, 299 (2006)
 
 \bibitem{Luzum:2010ag}
  M.~Luzum,
  [arXiv:1011.5173 [nucl-th]].
  
  \bibitem{Shen_priv}
  U.~Heinz and C.~Shen, private communication.
  
 \bibitem{Back:2004zg}
  B.~B.~Back {\it et al.} [ PHOBOS Collaboration ],
  Phys.\ Rev.\ Lett.\  {\bf 94}, 122303 (2005).
 
\bibitem{Busza:2007ke}
  W.~Busza,
  J.\ Phys.\ G {\bf 35}, 044040 (2008)

\bibitem{Kumar:2011de}
  L.~Kumar, f.~t.~S.~Collaboration,
  [arXiv:1101.4310 [nucl-ex]].
  
\bibitem{Niemi:2011ix}
  H.~Niemi, G.~S.~Denicol, P.~Huovinen {\it et al.},
  [arXiv:1101.2442 [nucl-th]].
  
\end{thebibliography}
\end{document}